\documentstyle[preprint,eqsecnum,aps]{revtex}
\begin{document}
\draft
\preprint{CALT-68-1958}
\title{Conservation Laws and Sum Rules in the Heavy Quark Limit}
\author{Chi-Keung Chow}
\address{California Institute of Technology, Pasadena, CA 91125}
\date{\today}
\maketitle
\begin{abstract}
In the heavy quark limit, hadrons appear as eigenstates of the light degrees
of freedom under the static color field of the heavy quark.
In this formalism, the weak form factors appear naturally as the overlaps of
the initial and final wavefunctions of the light degrees of freedom, and the
Bjorken and Voloshin sum rules are statements of conservation of probability
and energy.
Moreover, parity conservation can lead to a sum rule which relates weak form
factors at different kinematic points.
{}From this sum rule, model independent lower bounds on Isgur--Wise form
factors
can be obtained analytically.
\end{abstract}
\narrowtext
It has been realized that by taking the limit of QCD where the masses of heavy
quarks ($m_Q \gg \Lambda_{\rm QCD}$) go to infinity with their four-velocities
fixed, new symmetries not manifest in the full theory of QCD appear
\cite{1,13} .
In this limit, a heavy quark is a static source of color field which is
independent of its mass and spin.
It is exactly analogous to the situation in atomic physics, where the nucleus
is much heavier than the typical energy scale $\alpha^2 m_e$.
Then the nucleus will just be a static source of the electric field, and the
physics will be independent of its mass and spin.

Just like the electron cloud have different eigenstates in the electric field
of the nucleus, the ``brown muck'' around a heavy quark will also have
different eigenstates in the color field of the heavy quark.
Such eigenstates are, of course, the hadronic resonances.
Although we cannot solve QCD explicitly to predict the properties of these
hadronic resonances, we can make use of their completeness to construct sum
rules relating different transition matrix elements.

In this paper, we will follow this strategy and construct three different sum
rules, which correspond to the conversation laws of probability, energy and
parity respectively.
The first one will be seen to be just the well-known Bjorken sum rule
\cite{2,3,4}, while the second one is closely related to the Voloshin sum
rule \cite{5}.
Lastly, conservation of parity leads to a new sum rule relating weak form
factors at different kinematic points.
{}From this sum rule, model-independent lower bounds on Isgur--Wise form
factors
can be obtained analytically.

\section{The Bjorken Sum Rule}
To illustrate our formalism, we will consider heavy baryons containing a $b$
quark as our primary example.
In this case, $\Lambda_b$ is the ground state of the ``brown muck'' in the
color field of the $b$ quark.
When a $b$-quark in a $\Lambda_b$ with velocity $v$ decays into a $c$-quark
with velocity $v'$, all the ``brown muck'' notice is the change in velocity.
The ground state $|0\rangle$ in the color field of a heavy quark with velocity
$v$ is in general not an eigenstate of the color field of a quark with a
different velocity $v'$.
In the special case of $v=v'$, however, the color field is unchanged and the
``brown muck'' stay in the same state.
This gives the normalization of the Isgur--Wise form factors at the point of
zero recoil.

Denote the eigenstates of the ``brown muck'' in the color field of the charm
quark with velocity $v'$ by $|n'\rangle$, with $|0'\rangle$ the ground state
$\Lambda_c$.
The decay amplitude ${\cal M}(\Lambda_b(v)\rightarrow X^{n'}_c(v')e\bar\nu)$ is
proportional to the overlap of the initial and final states
$\varphi_{n'}(w)=\langle n'|0\rangle$, where $w=v\cdot v'$ and the ``brown
muck'' of $X^{n'}_c$ is in the state $|n'\rangle$.
This clearly hints that $\varphi_{n'}(w)$ are closely related to the weak form
factors, and indeed it is the case.
For example, consider the baryonic Isgur-Wise form factor $\eta(w)$, defined by
\begin{equation}
\langle \Lambda_c (v', s')|\bar c \Gamma b|\Lambda_b (v, s) \rangle
= \eta(w) \bar u(v', s') \Gamma u(v, s) .
\end{equation}
With $|\Lambda_b (v, s) \rangle=|b (v, s)\rangle \otimes |0\rangle$ and
$|\Lambda_c (v', s') \rangle=|c (v', s')\rangle \otimes |0'\rangle$, we have
\begin{eqnarray}
\langle \Lambda_c (v', s')|\bar c \Gamma b|\Lambda_b (v, s) \rangle
&=& \langle 0'|0\rangle \bar u(v', s') \Gamma u(v, s) \nonumber\\
&=& \varphi_{0'}(w) \bar u(v', s') \Gamma u(v, s) ,
\end{eqnarray}
we immediately have
\begin{equation}
\varphi_{0'}(w) = \eta(w) .
\label{bfr}
\end{equation}
Weak form factor of excited baryons can also be obtained in a similar way.
In particular, for $P$-wave baryons ,
\begin{equation}
\varphi_{n'}(w) = (w+1)^{1/2}\sigma^{n'}(w) .
\end{equation}
Where $\sigma^{n'}(w)$ is defined in Ref. \cite{3}.
In general, for a state with orbital angular momentum $l>0$, $\varphi_{n'}(w)
\sim |v'-v|^l \sim (w-1)^{l/2}$.
For $l=0$,  $\varphi_{n'}(w) \sim (w-1)^0$ for the ground state and $\sim
(w-1)^1$ for excited states.
This determines the behavior of $\varphi$ near the point of zero recoil.

By completeness of $|v'; n' \rangle$ we have the sum rule
\begin{equation}
|\varphi_{n'}(w)|^2 = \sum_{n'}\langle 0|n'\rangle\langle n'|0\rangle
= \langle 0|0\rangle = 1,
\label{psr}
\end{equation}
which states the conservation of probability.
We will see that this is equivalent to the Bjorken sum rule.

Traditionally \cite{3,4}, the Bjorken sum rule is based on consideration the
four-point function of $b$, $\bar b \Gamma c$, $\bar c \Gamma b$ and $\bar b$,
where $\Gamma = \gamma^\mu (1-\gamma_5)$.
The four-point function may be evaluated in quark language by perturbative QCD,
or in hadron language in the framework of heavy quark symmetry.
By duality the two results should be equal.
Hence,
\begin{equation}
h^{b \rightarrow c} (w) = \sum_{n'} h^{n'}(w)
\label{dual}
\end{equation}
where
\begin{equation}
h^{n'} (w)= \sum_{s'} \langle \Lambda_b (v)|\bar b \Gamma c|X^{n'}_c (v',s')
\rangle \langle X^{n'}_c (v',s')|\bar c \Gamma b| \Lambda_b(v)\rangle ,
\end{equation}
where $X^{n'}$ are multiplets in the heavy quark symmetry, and
\begin{equation}
h^{b \rightarrow c} (w) = \sum_{s,s'} \langle b (v,s)|\bar b \Gamma c|c
(v',s')\rangle \langle c (v',s')|\bar c \Gamma b|b (v,s)\rangle .
\end{equation}
Since the $\bar b \Gamma c$ and $\bar c \Gamma b$ in the definition of $h^{n'}$
act on heavy quarks but not on the light degrees of freedom, we expect $h^{n'}$
can be factorized into contributions from the heavy quark sector and those from
the light degrees of freedom.
The heavy quark sector will just reproduce $h^{b \rightarrow c}$, while the
contribution from the light degrees of freedom can be expressed in terms of
$\varphi_{n'}$.
Hence we end up with
\begin{equation}
h^{n'}(w) = h^{b \rightarrow c}(w) \langle 0|n' \rangle \langle n'|0 \rangle
= h^{Q_i \rightarrow Q_j}(w) |\varphi_{n'} (w)|^2
\end{equation}
After summation over all $n'$ and canceling the common factor of $h^{Q_i
\rightarrow Q_j}$ off both sides of Eq. (\ref{dual}), we end up with
\begin{equation}
1 = \sum_{n'}| \varphi_{n'} (w)|^2
\end{equation}
which is just Eq. (\ref{psr}) reproduced.
Replacing $\varphi$'s with the weak form factor, the equation becomes
\begin{equation}
1 = |\eta(w)|^2 + (w^2-1) \sum_q |\sigma^{(q)}(w)|^2  + {\cal O}^2 (w-1)
\label{bbsr},
\end{equation}
which is just the usual Bjorken sum rule.

Expanding about the point of zero recoil, the Bjorken sum rule can be
simplified to
\begin{equation}
{\bar \rho}^2 = \sum_q |\sigma^{(q)}(1)|^2
\label{bbdr}
\end{equation}
where the charge radius $\bar \rho$ of $\eta(w)$ is defined by
\begin{equation}
\eta(w) = 1 - {\bar \rho}^2(w-1) + ...
\end{equation}

Similar analysis can be made for the meson sector.
For a $b$-quark, the ground states with $s_\ell = {\textstyle {1\over2}}$ are
the $B$ and $B^*$ mesons.
The formulas are more complicated, however, as the mesonic Isgur--Wise form
factor $\xi(w)$ is not exactly $\varphi_0'(w)$.
In fact, it turns out that
\begin{equation}
|\varphi_{0'} (w)|^2 = \left({w+1 \over 2}\right)|\xi(w)|^2 .
\label{mfr}
\end{equation}
Putting $\varphi_{0'}(w)= 1$ recovers the Bjorken--Suzuki upper bound of
$\xi(w)$
\cite{2,15}.
\begin{equation}
\xi(w) \leq \Biggl({2\over1+w}\Biggr)^{1/2} .
\label{bb}
\end{equation}
For the $P$-wave excited mesons,
\begin{mathletters}
\begin{equation}
|\varphi^{(q)} ({\textstyle {1\over2}}^+; w)|^2 = 2(w-1)
|\tau_{1/2}^{(q)}(w)|^2,
\end{equation}
\begin{equation}
|\varphi^{(r)} ({\textstyle {3\over2}}^+; w)|^2
= (w-1)(w+1)^2 |\tau_{3/2}^{(r)}(w)|^2.
\end{equation}
\end{mathletters}
where the $\tau$'s are defined in the Ref. \cite{4}.
In this case, Eq.(\ref{psr}) becomes
\begin{equation}
1=\left({w+1 \over 2}\right)|\xi(w)|^2+(w-1) \Biggl[2 \sum_q
|\tau_{1/2}^{(q)}(w)|^2 + (w+1)^2 \sum_r |\tau_{3/2}^{(r)}(w)|^2 \Biggr] +
{\cal O}^2 (w-1)
\end{equation}

Defining the charge radius $\rho$ of $\xi(w)$ by
\begin{equation}
\xi(w) = 1 - \rho^2(w-1) + ...
\end{equation}
and the Bjorken sum rule is simplified to
\begin{equation}
\rho^2 = {\textstyle {1\over4}} + \sum_q |\tau_{1/2}^{(q)}(1)|^2
+ 2\sum_r |\tau_{3/2}^{(r)}(1)|^2 .
\label{mbdr}
\end{equation}
The extra $1\over4$ in Eq. (\ref{mbdr}) when compared to Eq. (\ref{bbdr})
is intriguing.
In this formalism it is clear that the $1\over4$ results from the
``uncanonical'' definition of $\xi(w)$ \cite{13}.
If the factor of $w+1\over2$ in Eq. (\ref{mfr}) is absorbed into the
definition of $\xi(w)$, the equation will have the same form as Eq.
(\ref{bfr}), and the $ 1\over4$ will not appear in the expansion.

\section{The Voloshin Sum Rule}
Returning to Eq. (\ref{psr}), it is noted that a more general sum rule holds
for an arbitrary operator ${\bf X}$:
\begin{equation}
\sum_{n'} \langle 0|{\bf X}|n'\rangle \langle n'|0\rangle
= \langle 0|{\bf X}|0\rangle .
\label{msr}
\end{equation}
In particular, if we put ${\bf X} = \openone$ in Eq. (\ref{msr}), Eq.
(\ref{psr}) is recovered.

Another case of interest is when ${\bf X} = H'$ the Hamiltonian in the color
field of a heavy quark with velocity $v'$.
Without loss of generality we choose the {\it final} velocity $v'=(1,{\bf 0})$.
Then $E_{n'}=\Delta m_{n'}=m_{X^{n'}_c}-m_{c}$ are just the excitation energies
of the resonances $X^{n'}_c$.
For the ground state, $\Delta m_{0}=\Lambda=m_D-m_c$ in the meson sector and
$\bar\Lambda=m_{\Lambda_c}-m_c$ in the baryon sector.
The right-handed side $\langle 0|H'|0\rangle$ is the energy expectation for a
moving ground state ``brown muck'' under the color field of a stationary heavy
quark of velocity $v'$.
By dimensional analysis we know that
\begin{equation}
\langle 0|H'|0\rangle = \Delta m_0 k(w),
\end{equation}
where $k(w)$ is a kinematic factor which depends on $w$ only.
Hence the whole sum rule reads as
\begin{equation}
\sum_{n'}\Delta m_{n'}|\varphi_{n'}(w)|^2=\Delta m_0 k(w) .
\label{esr}
\end{equation}

The functional form of $k(w)$ can be obtained in some definite scenarios.
For example, if we can regard $H'$ as completely kinetic, then $k(w)$ is just
the Lorentzian boost factor $\gamma$, which is just $w$.
A more realistic scenario sees $H'$ with two parts: the mass of the ``brown
muck'' in $v'$ frame and the potential energy of the ``brown muck'' in the
color field of a heavy quark with velocity $v'$.
If we ignore the quantum fluctuation of the color field, i.e, the heavy quark
is a classical, static source, we can choose the color potential of the heavy
quark such that
\begin{equation}
A^a_\mu ({\bf x}) = A^a(r)v'_\mu .
\end{equation}
The potential energy is of the form
\begin{equation}
U=\int d^3{\bf x}\; j^{a\mu} ({\bf x}) A^a_\mu ({\bf x}) ,
\end{equation}
where $j^{a\mu} ({\bf x})$ is the color current density of the ``brown muck''
and $a$ being the $SU(3)$ index.
By symmetry we have,
\begin{equation}
j^{a\mu} ({\bf x}) = j^a(r)v^\mu + t^\mu(r)
\end{equation}
where $t^\mu(r)$, the component of $j^{a\mu} ({\bf x})$ transverse to $v^\mu$,
is radially symmetric.
On integration, this transverse term vanishes by radial symmetry, and
\begin{equation}
U=w\int d^3{\bf x}\; j^a (r) A^a (r) .
\end{equation}
Hence when the ``brown muck'' is boosted from $v'$ to $v$, the potential energy
is increased by just the Lorentzian factor $w$.
Since the mass of the ``brown muck'' also increases by the same factor under
boost, we have $\langle 0|H'|0\rangle = \langle 0'|H'|0'\rangle w$, i.e, $k(w)
=w$ if we assume the color field of the heavy quark is purely classical.
It is probable that the statement is still valid if we take into account the
effects due to quantum fluctuation, though the author has not yet succeeded in
proving it.

If we assume $k(w)=w$ then Eq. (\ref{esr}) becomes
\begin{equation}
\sum_{n'}\Delta m_{n'}|\varphi_{n'}(w)|^2=\Delta m_0w
\end{equation}
which can be recasted into
\begin{equation}
\Delta m_0 (w-1) = \sum_{n'} (\Delta m_{n'}-\Delta m_0) | \varphi_{n'} (w)|^2.
\end{equation}
The quantity $E_{n'}=\Delta m_{n'}-\Delta m_0$ is just the mass difference over
the ground state.
In particular it is zero for $n'=0$, i.e, the term proportional to the
Isgur--Wise form factors vanishes.

For the meson sector, $\Delta m_{0'}=\Lambda=m_B-m_b$.
Substituting in the weak form factors, we obtain
\begin{equation}
\Lambda (w-1) = \sum_q E_{1/2}^{(q)} 2(w-1)|\tau_{1/2}^{(q)}(1)|^2
+ \sum_r E_{3/2}^{(r)} (w-1)(w+1)^2|\tau_{3/2}^{(r)}(1)|^2 + {\cal O}^2(w-1) .
\end{equation}
Canceling $(w-1)$ off both sides, and putting $w=1$, the sum rule becomes
\begin{equation}
\Lambda = \sum_q 2E_{1/2}^{(q)} |\tau_{1/2}^{(q)}(1)|^2 + \sum_r 4E_{3/2}^{(r)}
\end{equation}
which is just the Voloshin sum rule derived in Ref. \cite{5}.
On the other hand, in the baryonic sector, $\Delta m_{0'}=\bar \Lambda
=m_{\Lambda_b}-m_b$, and the sum rule reads as
\begin{equation}
\bar \Lambda = \sum_q 2E_1^{(q)} |\sigma^{(q)}(1)|^2 .
\end{equation}
The authors are not aware of any previous appearance of this sum rule in the
literature, except for Ref. \cite{6}, where this sum rule is obtained in the
large $N_c$ limit.
In fact, the results of Ref. \cite{6} can be reproduced in our formalism by
choosing a definite potential energy function, namely the isotropic harmonic
potential $V(r) = {\textstyle {1\over2}} \kappa r^2$.

\section{The Parity Sum Rule}
In this section, we will consider the case when we put  ${\bf X}=P'$, the
parity operator in the $v'$ frame, into Eq. (\ref{msr}).
The left-handed side becomes $\sum (-1)^{\pi_{n'}}|\varphi_{n'}(w)|^2$; where
$\pi_{n'}$ are the intrinsic parities of $|n'\rangle$.
$P'|0\rangle$ up to a phase is a ground state ``brown muck'' in the color field
of a heavy quark with velocity $\bar v = (w,-w{\bf v})$; hence the right-handed
side of Eq. (\ref{msr}) becomes $\langle 0|P'|0\rangle=(-1)^{\pi_{0'}}
\varphi_{0'}(W)$, $W=v\cdot\bar v$.
As result, Eq. (\ref{msr}) is simplified to
\begin{equation}
\sum_{n'} (-1)^{\pi_{n'}-\pi_{0'}}|\varphi_{n'}(w)|^2=\varphi_{0'}(W) .
\end{equation}
This sum rule is remarkable in the sense that it relates form factors at two
different kinematic points, $w$ and $W$.

Together with the Bjorken sum rule Eq.(\ref{psr}) we have
\begin{equation}
2\sum_{n'}^+ |\varphi_{n'}(w)|^2-1=\varphi_{0'}(W) .
\end{equation}
The $+$ above the summation means that the sum runs over the states with the
same parity as the ground states only.
Denoting the contribution to the sum from excited states as $R(w)$, the sum
rule reads
\begin{equation}
2\varphi_{0'}^2(w)-1+R(w)=\varphi_{0'}(W) .
\label{ssr}
\end{equation}
Since $\varphi_{0'}(w)$ is the Isgur--Wise form factors up to possibly a known
kinematic factor, a bound on $R(w)$ may give a model-independent bound on the
Isgur--Wise form factors.

Since $R(w)$ is a sum of absolute squares, $R(w)\geq 0$, and
\begin{equation}
2\varphi_{0'}^2(w)-1\leq\varphi_{0'}(W) .
\label{lb1}
\end{equation}
We will change the independent variable from $w$ to the ``boost angle''
$\alpha$, which is related to $w$ by $w=\cosh (\alpha)$.
We will also change the dependent variable from $\varphi_{0'}(\alpha)$ to
$f(\alpha)$, which is related to $\varphi_{0'}(\alpha)$ by $\varphi_{0'}
(\alpha)=\cos(f(\alpha)\alpha)$.
This greatly simplifies the equation as
\begin{equation}
W=2w^2-1=\cosh (2\alpha)
\end{equation}
and
\begin{equation}
2\varphi_{0'}^2(\alpha)-1=2\cos^2(f(\alpha)\alpha)-1=\cos(2f(\alpha)\alpha) .
\end{equation}
Hence Eq. (\ref{lb1}) becomes
\begin{equation}
\cos(2f(\alpha)\alpha)\leq \cos(2f(2\alpha)\alpha) .
\label{lb2}
\end{equation}
We expect $\varphi_{0'}$ to be a decreasing function.
Hence Eq. (\ref{lb2}) implies
\begin{equation}
f(\alpha)\geq f(2\alpha) .
\end{equation}
Since Isgur--Wise form factors are continuous, this simply states that
$f(\alpha)$ is also a decreasing function.
Moreover, since $\varphi_{0'}$ is expected to be nodeless and approaches zero
as $w\rightarrow \infty$ ($\alpha\rightarrow\infty$), we have
\begin{equation}
f(\alpha)\leq{\pi\over 2\alpha} ;
\end{equation}
\begin{equation}
f(\alpha)\rightarrow{\pi\over 2\alpha} ,\;\;\alpha\rightarrow\infty.
\end{equation}
Finally, the boundary condition at $w=1$ ($\alpha=0$) can be given in terms of
the derivative of the Isgur--Wise form factor at the point of zero recoil.
\begin{equation}
f(\alpha=0)=\bar\rho .
\end{equation}
The unique maximal $f(\alpha)$ satisfying the conditions above is
\begin{equation}
f^{\rm max}(\alpha)=\cases{\bar\rho, &$\alpha<{\pi\over 2\bar\rho}$; \cr
{\pi\over 2\alpha}, &$\alpha>{\pi\over 2\bar\rho}$.  \cr}
\end{equation}
Putting into the original form of $\varphi_{0'}(w)$, a model-independent lower
bound for $\varphi_{0'}(w)$ can be obtained.
\begin{equation}
\varphi_{0'}^{\rm min}(w)=\cases{\cos(\bar\rho \cosh^{-1}(w)), &$w<\cosh(\pi/
2\bar\rho)$; \cr 0, &$w>\cosh(\pi/2\bar\rho)$.  \cr}
\end{equation}
This is a lower bound for all possible forms of $\varphi_{0'}(w)$ {\it with the
same $\bar\rho$}.

Since $\eta(w)=\varphi_{0'}(w)$ in the baryon sector, the lower bound above can
be applied to the baryon case directly.
Plots of $\eta^{\rm min}(w)$ for different $\bar\rho$ are shown in Fig. 1.
This lower bound rules out some particular forms of $\eta(w)$ like the
piecewise
linear model
\begin{equation}
\eta(w)=\cases{1-\bar\rho^2(w-1), &$w<1+\bar\rho^{-2}$; \cr
0, &$w>1+\bar\rho^{-2}$.  \cr}
\end{equation}
It is known that, in the large $N_c$ limit, the baryonic Isgur--Wise form
factor has an exponential form \cite{7,8},
\begin{eqnarray}
\eta(w)&=&\exp(-\bar\rho^2(w-1))\nonumber\\&=&1-\bar\rho^2(w-1)
+{\bar\rho^4\over2}(w-1)^2 + ......
\end{eqnarray}
while our lower bound, in a Taylor series, is
\begin{equation}
\eta^{\rm min}(w)=1-\bar\rho^2(w-1)+({\bar\rho^2\over6}+{\bar\rho^4\over6})
(w-1)^2 + ......
\end{equation}
In the large $N_c$ limit, $\bar\rho^2\sim N_c^{3/2}$ is large, and the bound is
satisfied.

On the other hand, the mesonic Isgur--Wise form factor $\xi(w)$ is given by the
slightly more complicated Eq. (\ref{mfr}).
Hence, the lower bound for $\xi(w)$ is
\widetext
\begin{equation}
\xi^{\rm min}(w)=\cases{\bigr({2\over 1+w}\bigr)^{1/2}\cos((\rho^2-{\textstyle
{1\over4}})^{1/2} \cosh^{-1}(w)), &$w<\cosh(\pi/2(\rho^2-{\textstyle
{1\over4}})^{1/2})$; \cr 0, &$w>\cosh(\pi/2(\rho^2-{\textstyle
{1\over4}})^{1/2})$.  \cr}
\end{equation}
Plots of $\xi^{\rm min}(w)$ for different $\bar\rho$ are shown in Fig. 2.

Expanding in a Taylor series, we have
\begin{equation}
\xi^{\rm min}(w)=1-\rho^2(w-1)+({\rho^2\over3}+{\rho^4\over6})(w-1)^2 + ......
\end{equation}
When compared with the ISGW \cite{9}, BSW \cite{10,11,12} and pole \cite{13}
parametrizations of the mesonic Isgur--Wise form factor,
\begin{eqnarray}
\xi_{\rm ISGW}(w)&=&\exp(-\rho_{\rm ISGW}^2(w-1))\nonumber\\
&=&1-\rho_{\rm ISGW}^2(w-1)+{\rho_{\rm ISGW}^4\over2}(w-1)^2 + ......,
\end{eqnarray}
\begin{eqnarray}
\xi_{\rm BSW}(w)&=&{2\over w+1} \exp\left((1-2\rho_{\rm BSW}^2)
{w-1\over w+1}\right) \nonumber\\&=&1-\rho_{\rm BSW}^2(w-1)+(-{1\over4}
+{\rho_{\rm BSW}^2\over2}+{\rho_{\rm BSW}^4\over8})(w-1)^2 + ......
\end{eqnarray}
\begin{eqnarray}
\xi_{\rm pole}(w)&=&\left({2\over w+1}\right)^{2\rho_{\rm pole}}\nonumber\\
&=&1-\rho_{\rm pole}^2(w-1)+({\rho_{\rm pole}^2\over4}+{\rho_{\rm pole}^4
\over2})(w-1)^2 + ......
\end{eqnarray}
\narrowtext
We found that, in order to satisfy the lower bound, we must have
$\rho_{\rm ISGW}\geq1$ and $\rho_{\rm pole}\geq0.5$.
$\xi_{\rm BSW}$, on the other hand, does not satisfy the bound for all values
of $\rho_{\rm BSW}$.

Last of all, it's also worth mentioning once more that the Bjorken--Suzuki
upper bound of $\xi(w)$, which is given in Eq. (\ref{bb}), can be recovered by
putting the obvious inequality $\varphi_{0'}(w)\leq1$ into Eq. (\ref {mfr}).
The counterpart of this upper bound in the baryon sector is the trivial
statement $\eta(w)\leq1$.

\bigskip
In the paper, we have discussed the Bjorken, Voloshin, and parity sum rules
within the same framework.
This is not meant to be a rigorous derivation (as in Ref. \cite{3,4,5}) but
just an intuitive picture making the relationship between the sum rules and the
conservation laws behind them more transparent.
We have treated the case for heavy mesons and the $\Lambda_Q$-type baryons, but
similar analysis can be made in the $\Sigma_Q$-type baryon sector \cite{8,14}
as well.

I must thank Mark Wise for bringing the Bjorken sum rule to my attention, and
Ming Lu for the Voloshin sum rule.
This work was supported in part by the U.S. Dept. of Energy under Grant No.
DE-FG03-92-ER 40701.

\begin{figure}
\caption{$\eta^{\rm min}(w)$ for different values of $\bar\rho$.
 From top to bottom $\bar\rho^2=0.25$, 0.50, 0.75, 1.00, 1.25 and 1.50.  }
\end{figure}
\begin{figure}
\caption{$\xi^{\rm min}(w)$ for different values of $\rho$.
 From top to bottom $\rho^2=0.25$, 0.50, 0.75, 1.00, 1.25 and 1.50.  }
\end{figure}
\end{document}